\documentstyle[prl,aps,epsf]{revtex}
\begin{document}
\draft
\tighten
\twocolumn[\hsize\textwidth\columnwidth\hsize\csname @twocolumnfalse\endcsname
\title{Tunneling edges at strong disorder}
\author{Jonathan Miller$^a$ and A.G.Rojo$^b$}
\address{
$a$ James Franck Institute, The University of Chicago,
Chicago, IL 60637}
\address{
$^b$The Harrison Randall Labroratory
of Physics, The University of Michigan, Ann Arbor, MI 48109-1120}
\maketitle

\begin{abstract}

Scattering between edge states that bound one-dimensional domains of
opposite potential or flux is studied, in the presence of strong
potential or flux disorder. A mobility edge is found as a function
of disorder and energy, and we have characterized the extended regime.
In the presence of flux and/or potential disorder, the localization
length scales exponentially with the width of the
barrier. We discuss implications for the random-flux problem.

\end{abstract}
\vskip2pc] \narrowtext

The remarkable transport properties of a 2d electron gas in
a strong perpendicular magnetic field are well-known\cite{prange}:
jumps in Hall resistance and peaks in longitudinal resistance
occur as the Fermi level crosses the single energy, at the center
of the Landau band, where extended states are believed to lie.

Experiments and numerical studies show that, as the Fermi level
$E_c + \delta E$ approaches the center of the band, $E_c$, the correlation
length $\zeta$ diverges: $\zeta \propto \vert \delta E \vert ^{-\nu_q}$, where
$\nu_q \approx 7/3$. Disorder plays a crucial role in all models
for this transition. In a strong magnetic field $B$, eigenstates are confined
to equal-energy contours of a disordered potential, slowly-varying
on scales of order the magnetic length $l_B=(\hbar c/e B)^{1/2}$.
For a generic random, symmetric potential $V(\vec{r})$ with characteristic
length\cite{isichenko} $\lambda_0$, each of these equipotentials,
for energies not too
close to $E_c= \langle V \rangle$, will traverse a region of finite size,
exceeding the cluster diameter $a(\vert \delta E \vert)$ in cross-section with
exponential rarity. When $\delta E \to 0$,
long-range spatial correlations emerge in the equipotentials as
the cluster diameter diverges at the  continuum percolation
transition\cite{isichenko}.

A model for this transition was proposed by Trugman\cite{trugman},
who argued that one might neglect tunneling among
distinct equipotentials because of the small overlap of their
respective wave-functions. If this omission were valid, then
the localization length would diverge as the spatial extent
of the percolating eigenfunctions. The correlation exponent $\nu_c$
for the classical percolation transition is known to be $4/3$.

Mil'nikov and Sokolov\cite{milnikov} [MS] attempted to calculate
the correction to Trugman's classical picture arising from quantum
tunneling between equipotentials. They argued that the dominant
contribution of tunneling derived from the saddle-points of the
potential along the largest eigenstates. Since these saddle-points
occur between longest equipotentials that traverse distances of
order $a(\vert \delta E \vert) = \vert \delta E \vert^{-4/3}$,
and the correction they obtained from tunneling was inversely
proportional to $\vert \delta E \vert$, MS suggested that $7/3 = 4/3 + 1$.

Their proposal, even were it correct, is not supported by their argument.
While saddle-points between longest equipotentials are rare, each
equipotential is also a (bond) percolation hull, bounding a complex, fractal
network of wave-functions with energy $E_c$. MS offered no justification
for their neglect of the scattering between the hull and this internal
network, which could lead to localization via back-scattering.

The most widely accepted model for the Hall transition, the
Chalker network model\cite{lee_chalker}, yields numerically a $\nu_q$
close to $7/3$.
The Chalker model consists of a regular lattice of tunneling
saddle-points, to be contrasted with the \lq topologically disordered\rq\
network of equi-contours that underlies the classical percolation
transition. This contrast with classical percolation has lead
to the claim that quantum tunneling fundamentally alters the classical
transition, and that the relation $\nu_q  \approx 1 + \nu_c$ is
merely coincidence.

This issue assumes special
importance for understanding eigenstates in a random--flux
background (with vanishing
mean flux). If quasi-classical arguments applied, one would expect
an extended state in the presence of random--flux that exhibited
much the same properties as obtained for potential disorder;
however, with appropriate modification to incorporate the
vanishing net chirality of the random--flux problem,
the Chalker model displays no extended states\cite{lee_chalker}.

Thus, the Chalker model appears to suggest that a picture of
of wave-functions, whose spatial extent is dominated by classical
equipotentials, is naive. This state of affairs prompted
us to reconsider tunneling among distinct equipotentials. In
particular, we would like to understand when it may be possible
to think of equipotentials as distinct classical units, and to
calculate transport properties of a network of eigenstates
by perturbing around their spatial configurations.

Such an understanding might be relevant to a variety of physical
systems: not only to the QHE and the random-flux problem,
but also to gauge-field theories of flux phases, where boundaries between
distinct flux domains create a network of one-dimensional
edges\cite{chklovskii}. Edges created by nonhomogeneous magnetic
field in the
{\it absence} of disorder\cite{peeters}, or in the context of
the quantum Hall effect with small disorder\cite{chklovskii}
have been studied elsewhere; here
we study tunneling between equipotentials
in the presence of {\it strong} potential and/or flux disorder.

We construct a pair of one-dimensional wave-functions
on a strip of length $L$ and width $M$ lattice constants $a$.
Our lattice Hamiltonian is:
$$
	H = \sum_{i}V_i \vert i \rangle \langle i \vert
	+
	\lbrace
	    \sum_{i,j}t_{ij} \vert i \rangle \langle j \vert + \hbox{$h.c.$}
	\rbrace
	\eqno\hbox{(1)}
$$
where $V_i$ denotes the on-site energy, and the hopping
matrix elements $t_{ij}$ vanish except for nearest neighbors.
Finite magnetic flux is included by allowing complex $t_{ij}$,
where the local phase of $t_{ij}$ is determined by the flux enclosed within
a square plaquette of the lattice. (Energies are
in units of $\vert t \vert =1$, and we also take the lattice
constant $a=1$).  We calculate localization
lengths for this system numerically by computing the largest
eigenvalue of a product of $L$ transfer matrices, each of which
adds a single row of $M$ lattice sites at the end of the strip,
and finite-size scaling\cite{mackinnon}. These methods
are now standard, although to our knowledge they have not
been applied before to the geometries we discuss here.

A pair of parallel edges, separated by roughly $M/2$
lattice sites (one edge centered between sites $M/2$ and $M/2+1$, the
other between sites $M$ and $1$) was constructed in several ways
($i$ denotes transverse coordinate on the strip):

\vskip0.25pc
($1a$) Two parallel stripes of flux $B_0$ with equal magnitude and
opposite sign:
$B_i = -B_0 \theta(M/2-i) + B_0 \theta(i-M/2)$,
random potential $V$ is chosen on the interval
$[-W/2,W/2]$.

($1b$) Two parallel stripes of uniform potential $V_0$ and
width $M/2$, with equal magnitude and opposite sign in uniform flux
$B_0$. Superposed on the potential stripes is random potential,
chosen from $[-W/2,W/2]$.
\vskip0.25pc

($2a$) Two parallel stripes of flux, with flux through
plaquette $i$ chosen randomly from $[0,B_0]$ for $i \le M/2$,
and from $[-B_0,0]$ for $i > M/2$.

($2b$) Two parallel stripes of uniform potential $V_0$
and width equal in magnitude and opposite in sign. Superposed
on the potential stripes was random flux with finite mean $B_0$,
chosen from the interval $[B_0-\delta B,B_0+\delta B]$;
$\vert \delta B \vert < \vert B_0 \vert$
\vskip0.25pc

All random quantities were chosen uniformly on the specified
intervals, and independently on each lattice site.
Periodic boundary conditions were imposed, except where
explicitly noted below. We have concentrated in the regime of
strong disorder, with $W$ on order of the inter-Landau subband spacing.

Flux and potential configurations differing locally from
those listed above
over distances of a few lattice constants did not affect the
scalings we discuss below. We also allowed our edges
to meander randomly in the $y$ direction, and their
transverse separation to fluctuate. Provided that the
$y$ coordinate of an edge was allowed to vary by
at most a few lattice constants for each lattice site traversed
in the longitudinal direction, our scalings similarly
remained unchanged from those we found for configurations.
We emphasize that our results are
insensitive to details. The features of our edges that
are important for the following discussion appear to be
their low curvature, and their local (mean) flux/potential gradient,
which remains constant as we increase the width of the barriers.

We studied the dependence of the longitudinal
localization length upon $S$, the width of the stripes
in units of the lattice constant. Although we have for
simplicity described geometries with two stripes only ($S=M/2$),
our results are unchanged for multiple pairs of stripes.

\begin{figure}
\epsfysize=5.0in
\vspace{5pt}
\caption{
Localization length versus energy for $W=5.0$ ($a$) and $W=5.5$ ($b$), and
different
 values of strip width;
 $M=8$ (open circles), $16$ (full circles), $32$(squares), $64$ (diamonds).
 The localization length $\lambda_M$ has a relative error of $2\%$}
 \end{figure}
Figure $1(a)$ displays, for geometry (1a), $\lambda_M$ as
a function of $E$ and stripe width. The intersections
of the curves indicate mobility edges: in the localized
regime, $\lambda_M \propto \lambda_\infty$, a constant.
     For $B_0= \pi/4$, this mobility edge disappears
     as indicated by figure 1(b) when the potential disorder $W$
     exceeds roughly $W_{\rm cr}\sim  5.5$, yielding localization at all
energies.
Figure $1(a)$ suggests that in the extended regime,
$\lambda_M \propto M^\alpha$ with $\alpha \simeq 2$;
however, we found that for larger $M$,
$\lambda_M$ increases exponentially, scaling as
 $\lambda_M \propto M^\alpha
\exp(M/\Lambda)$, where $\Lambda$
 represents a length scale determined by properties of
the bulk. This exponential increase is apparent also for smaller
values of disorder ($W\ll W_{\rm cr}$). Scaled
curves displaying this behavior are shown in figure 2.

The critical properties of this transition will
be discussed elsewhere\cite{rojo}; our present concern is the
scaling with stripe width in the extended regime only.

\begin{figure}
\caption{Scaled localization length in the extended regime for $W=2.0$, and
different energies: $E=1.45$ (full squares), $3.$ (open circles),
$1.475$ (diamonds), $1.5$ (triangles), $2.8$ (open squares), $2.6$ (crosses).
Curves were collapsed by successive displacement along the horizontal axis;
at each step the interpolated (not extrapolated) displacements were
minimized by least--squares. }
\end{figure}
The $M^2$ scaling we find for $M < \Lambda $ differs from
that obtained for extended {\it bulk} states in potential
disorder and uniform flux with periodic boundaries, where
$\lambda_M \propto M$
is expected at a band center, and is asymptotically constant
elsewhere. (This latter geometry, in contrast to the geometries
studied here, is isotropic). A $\lambda_M/M$ increasing with $M$
above the band center has been observed previously for uniform
flux, Dirichlet boundary conditions, and potential disorder;
however, those studies attempted no quantitative analysis of this
phenomenon\cite{johnston}. We have found numerically that this
geometry also yields $M^\alpha \exp (M/\Lambda)$ scaling, and that
this behavior is consistent with the earlier published data\cite{johnston}.

We obtained this exponential scaling for all edge configurations we
studied, including $(1a,b)$ and $(2a,b)$. $V_0$ was set to $1.0$
or $2.0$ in cases $(1b),(2b)$; the values of $B_0$ and $V_0$
were unimportant provided that the Landau subbands were not
separated by much less than the characteristic scale of variation
of $W$.

When boundary conditions of the form $1a$--$2b$ or Dirichlet are
imposed, chiral states traverse the boundaries. In the absence
of edge--edge scattering, extended
states propagate along these edges; the only mechanism for
localization of the edge wave-functions is scattering with states
of opposite chirality at the other edge. We now argue that this
inter--edge hopping gives rise to the  scaling we have observed.

At the quasi--classical level, an edge state lives on a line of
constant potential. For strong disorder, an equipotential
can wander into the bulk and meet the opposite edge, yielding
backscattering. On an infinite plane, these equipotentials
(percolation hulls) circumscribe islands of finite extent,
the typical size given by $\delta E ^\nu$, $\delta E$ the
energy of the state with respect to the center of a magnetic subband.
At the center of a subband, there exists an infinite equipotential
that percolates through the sample. In this picture, the edge--to--edge
scattering will depend on the probability of finding islands of size
larger than $S$. The exponential rarity
of these islands, sufficiently far from the percolation threshold,
accounts for the exponential increase in localization length.
If the typical island size $\ell$ is $\ell \gg  S$, then we expect
$\lambda_e$ to display algebraic behavior.

When the disorder is small enough, $ W \ll \Delta $, $\Delta$
the gap between magnetic subbands, edge states of energy $E_0$
sufficiently removed from a magnetic subband center $E_c$
may not be connected to one another through equipotential
islands. Thus if $\vert E_0 -E_c \vert >W$,
our quasi-classical
argument that neglects tunneling among distinct equipotentials
implies a divergent localization length for finite edge
separation $S$. The inclusion of quantum tunneling would restore
the finite value of the localization length expected at finite $S$,
but the spatial attenuation of quantum scattering between edge
states as a function of their separation $S$ is then governed by
the magnetic length. The resulting localization lengths
are of a completely different order of magnitude than those
we find for edge states connected by the disorder-generated
quasi--classical equipotentials. Indeed, for sufficiently small
$W$, as the Fermi energy $E_0$ crosses into the gap $\vert E_0-E_c \vert >W$
the computed localization length increases abruptly by many
orders of magnitude, to values beyond our numerical control.

Lee and Chalker\cite{lee_chalker} have proposed a variant of the
original one-channel network model to study localization in
a random-flux background. In order to maintain a vanishing
$\sigma_{xy}$, they stipulated that each node describe equivalent
tunneling for the two channels of {\it opposite} chirality, so that
the node is represented by scattering parameters $\theta_1,\theta_2$
satisfying
$\sinh(\theta_1)=1/\sinh(\theta_2)$.
They obtain no extended state in this model.

Within the framework of edge--edge scattering described in this
paper, their conclusion
is readily understood: no configuration of their nodes can be
equivalent to a pair of edge states on opposite sides of a long
flux barrier. Rather, backscattering can occur at every node
with amplitude equal to the forward scattering that occurs at
that node. In contrast, our backscattering amplitude is exponentially
less than our forward scattering amplitude, for the entire length
of the flux barrier. In a random background, this length is set by
the quenched flux disorder, i.e. the classical path.

This overall picture supports the general notion, mentioned
in our introduction, that the classical paths (equipotentials)
make up the dominant contribution to transport, and that
tunneling among equipotentials may be viewed as a perturbation
of the classical background

This picture is also consistent with unpublished numerical studies of
one-dimensional defects in Chalker lattices\cite{miller}. In Chalker
networks, local isotropy is maintained by an antiferromagnetic choice
of node parameter $\theta$ on alternate sublattices. One can create
edges in a network by introducing defects in the antiferromagnetic
ordering, forming domain boundaries across which the sublattice
alternation is displaced by a half unit--cell. It is found
that pairs of linear defects (comparable to the edges we study here)
in a {\it one-channel} Chalker model display an exponential increase
of localization length with increasing separation. {\it Two-channel}
Chalker models display a richer behavior. Nodes of the form
described above ($\sinh(\theta_1)=1/\sinh(\theta_2)$) never exhibit
an extended regime, whereas nodes satisfying $\theta=\theta_1=\theta_2$
with $\sinh(\theta)=s,1/s$ on the alternate sublattices\cite{zhang} show
decreasing localization lengths as a function of system width $M$ for
$s$ close enough to $1$, and exponentially increasing localization lengths
as a function of $M$ for $s \geq s_c > 1$. Details will be given
elsewhere\cite{miller}.

In summary, we have described numerical studies of tunneling
between one-dimensional edge states created by boundaries between
regions of differing potential and/or flux. These systems display
distinct regimes as a function of energy and disorder;
we have characterized the extended regime, in which localization
lengths increase exponentially with the spatial separation of the
edges, and we have offered a quasi--classical explanation of this
behavior. The fundamental contribution of the classical paths,
whether formed by magnetic-flux or potential edges,
suggests to us that extended states ought to exist in a random--flux
background.

%
\acknowledgments
J.M. would like to thank D.P. Arovas, L.P. Kadanoff, and S.C. Zhang
for advice and discussions.
This work was supported in part by the MRSEC Program of the
National Science Foundation under Award Number DMR-9400379.
We also thank the National Center for Supercomputer Applications (NCSA)
(Grant No. DMR930021N) for access to NCSA Cray Y-MP supercomputer.

\end{document}